\documentclass{elsart}
\usepackage{graphicx}
\usepackage{amsmath}
\usepackage{amssymb}

\begin{document}

\begin{frontmatter}

\title{Nonequilibrium phenomena in adjacent electrically isolated nanostructures}

\author[address1,address2]{V.S.~Khrapai\thanksref{thank1}}
\author[address1]{S.~Ludwig}
\author[address1]{J.P.~Kotthaus}
\author[address3]{H.P.~Tranitz}
\author[address3]{W.~Wegscheider}

\address[address1]{Center for NanoScience and Department
f$\ddot{\text{u}}$r Physik,
Ludwig-Maximilians-Universit$\ddot{\text{a}}$t,
Geschwister-Scholl-Platz 1, D-80539 M$\ddot{\text{u}}$nchen,
Germany}

\address[address2]{Institute of Solid State Physics RAS,
Chernogolovka, 142432, Russian Federation}
\address[address3]{Institut f$\ddot{\text{u}}$r Experimentelle und
Angewandte Physik, Universit$\ddot{\text{a}}$t Regensburg, D-93040
Regensburg, Germany}

\thanks[thank1]{Corresponding author. E-mail: dick@issp.ac.ru;\\ Fax: +7-496-524-9701}

\begin{abstract}
We report on nonequilibrium interaction phenomena between adjacent
but electrostatically separated nanostructures in GaAs. A current
flowing in one externally biased nanostructure causes an
excitation of electrons in a circuit of a second nanostructure. As
a result we observe a dc current generated in the unbiased second
nanostructure. The results can be qualitatively explained in terms
of acoustic phonon based energy transfer between the two mutually
isolated circuits.
\end{abstract}
\end{frontmatter}

\section{Introduction}

Present lithography allows the fabrication of adjacent
nanostructures with a spatial separation of order 100 nm. Weak
Coulomb interaction between the electrons of two neighboring
nanostructures is widely used in charge detection schemes (see
e.g. Ref.~\cite{marcus}). Coulomb interaction between adjacent
one-dimensional (1D) wires can give rise to a frictional current
drag~\cite{zverev}. In finite magnetic fields the observation of
opposite signs of the drive and drag currents has been interpreted
in terms of a negative Coulomb drag~\cite{tarucha}. However, it
remains an open question whether the Coulomb interaction provides
the only relevant interaction mechanism in the low temperature
limit.

Here we report on novel nonequilibrium phenomena in adjacent but
electrostatically separated nanostructures, laterally defined in a
two-dimensional electron gas (2DEG) of a GaAs/AlGaAs
heterostructure. Thanks to the electrical separation, it is
possible to apply an arbitrary dc source-drain bias (hence a dc
drive current) to a drive-nanostructure while maintaining a second
detector-nanostructure at a small or zero bias. We observe that
the drive current can generate a finite dc current in the isolated
detector-nanostructure. The results are qualitatively explained in
terms of excitation of electrons in the detector circuit via
acoustic phonon based energy transfer from the drive circuit.

\section{Experimental details}

\begin{figure}[t]
\begin{center}
\includegraphics[clip,width=0.4\linewidth]{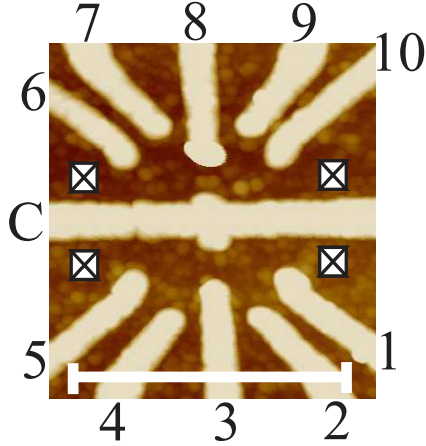} \caption{AFM micrograph of the nanostructure.
Metal gates on the surface of the heterostructure are shown in
bright tone. Crossed squares mark contacted 2DEG regions. The
scale bar equals 1~$\mu$m.} \label{fig1}
\end{center}
\end{figure}

The GaAs/AlGaAs heterostructure used for e-beam lithography
contains a 2DEG $90$~nm below the surface, with an electron
density of $n_S=2.8\times10^{11}$~cm$^{-2}$ and a low-temperature
mobility of $\mu=1.4\times10^6$~cm$^2/$Vs. The sample layout is
shown in fig.~\ref{fig1}. The (upper) drive-nanostructure contains
a quantum point contact (QPC) defined by gates 8 and C and
referred to as a drive-QPC. Throughout the paper the drive-QPC
conductance is tuned to nearly half a conductance quantum
$G_\text{drive}\approx e^2/h$. The (lower) detector-nanostructure
is separated by gate C and can be defined by gates 1 to 5. The dc
current measurements are performed in a dilution refrigerator at a
temperature of the electron system below 150~mK. In both circuits,
a positive sign of the current corresponds to electrons flowing to
the left. Below we present the results for a double quantum dot
(DQD) or a QPC used as the detector-nanostructure.

\section{Double-dot quantum ratchet}

\begin{figure}[t]
\begin{center}
\includegraphics[clip,width=0.8\linewidth]{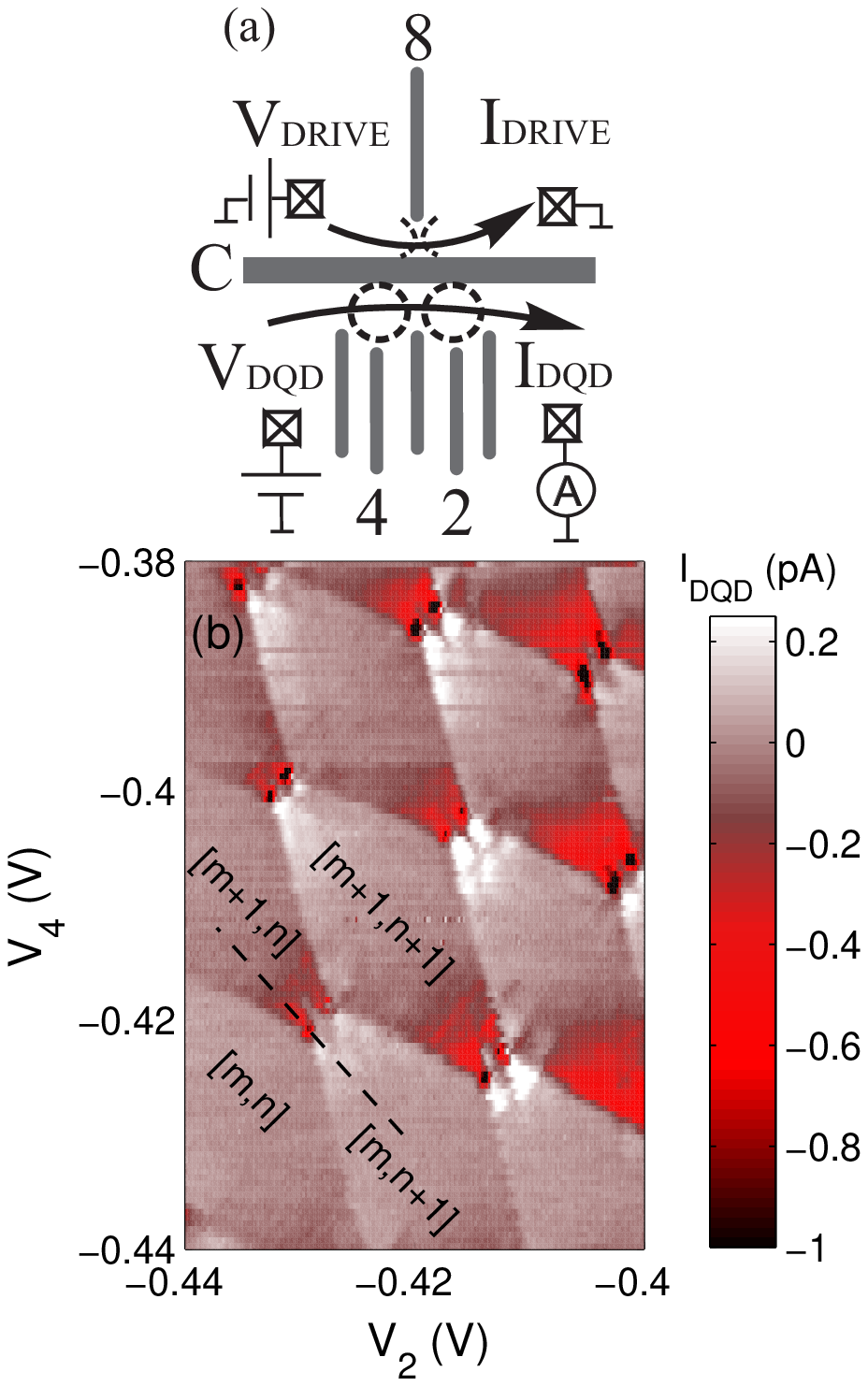}
 \caption{(a) -- Sketch of the measurement with a DQD-detector.
(b) -- Grey-scale plot (color online) of $I_\text{DQD}$ on the DQD
stability diagram. The data are taken at fixed
$V_\text{DRIVE}=-1.45$~mV and $V_\text{DQD}\approx-20\mu$V. The
charge configurations of the DQD for four regions of the stability
diagram are marked by two numbers as described in the text. The
dashed line marks the trace through one of the triple points along
which the data of Fig.~\ref{fig3} were taken.} \label{fig2}
\end{center}
\end{figure}

A schematic measurement layout for the case of the DQD-detector is
shown in Fig.~\ref{fig2}a. The DQD is formed by negatively biasing
gates 1 to 5 and represents two quantum dots tunnel-coupled in
series. The electron occupancies of the right and left dots are
mainly controlled by voltages applied to gates 2 and 4,
respectively. Both dots have single-particle level spacings of
about 100~$\mu$eV and charging energies of about 1.5~meV. The DQD
is tuned in a weak coupling regime, i.e. the interdot tunnel
splitting ($t\sim0.1~\mu$eV) is much smaller than the tunnel
coupling of the dots to the respective leads
($\Gamma\sim40~\mu$eV). A fixed small source-drain bias of
$V_\text{DQD}\approx-20\mu$V is applied across the DQD and a
current $I_\text{DQD}$ is measured with a current-voltage
amplifier.

In Fig.~\ref{fig2}b we show a color-scale plot of $I_\text{DQD}$
as a function of gate voltages $V_2,V_4$ controlling the charge
configuration of the DQD in presence of a relatively high
drive-QPC bias $V_\text{DRIVE}=-1.45$~mV. The hexagon-shaped
regions of the charging diagram are the regions of fixed ground
state charge configurations. Each configuration is referred to by
two numbers [m,n] corresponding to m (n) electrons occupying the
left (right) quantum dot (Fig.~\ref{fig2}b). The boundaries
between the neighboring ground state configurations are seen as
two sets of straight lines with different
slopes~\cite{vanderwiel}. Similar to the case of a conventional
DQD conductance measurement, at the intersections of the lines
(triple points) the chemical potentials of both quantum dots are
equal and within the DQD bias window. Here sharp resonance
tunnelling peaks are seen (black points in Fig.~\ref{fig2}b).
However, in contrast to conventional measurements, in the presence
of strong enough drive bias ($V_\text{DRIVE}\gtrsim1$~mV) finite
current is also observed away from the triple points in the regime
of ground state Coulomb blockade. The sign of $I_\text{DQD}$
depends on the position on the charging diagram of
Fig.~\ref{fig2}b. Below and to the right hand side from the triple
points $I_\text{DQD}>0$, whereas above and to the left hand side
$I_\text{DQD}<0$ (respectively, bright and dark tones of the
color-scale). Note that it is the abrupt change of the generated
$I_\text{DQD}$ with the change of the DQD ground state
configuration which makes the boundaries of the charging diagram
visible in Fig.~\ref{fig2}b.
\begin{figure}[t]
\begin{center}
\includegraphics[clip,width=0.6\linewidth]{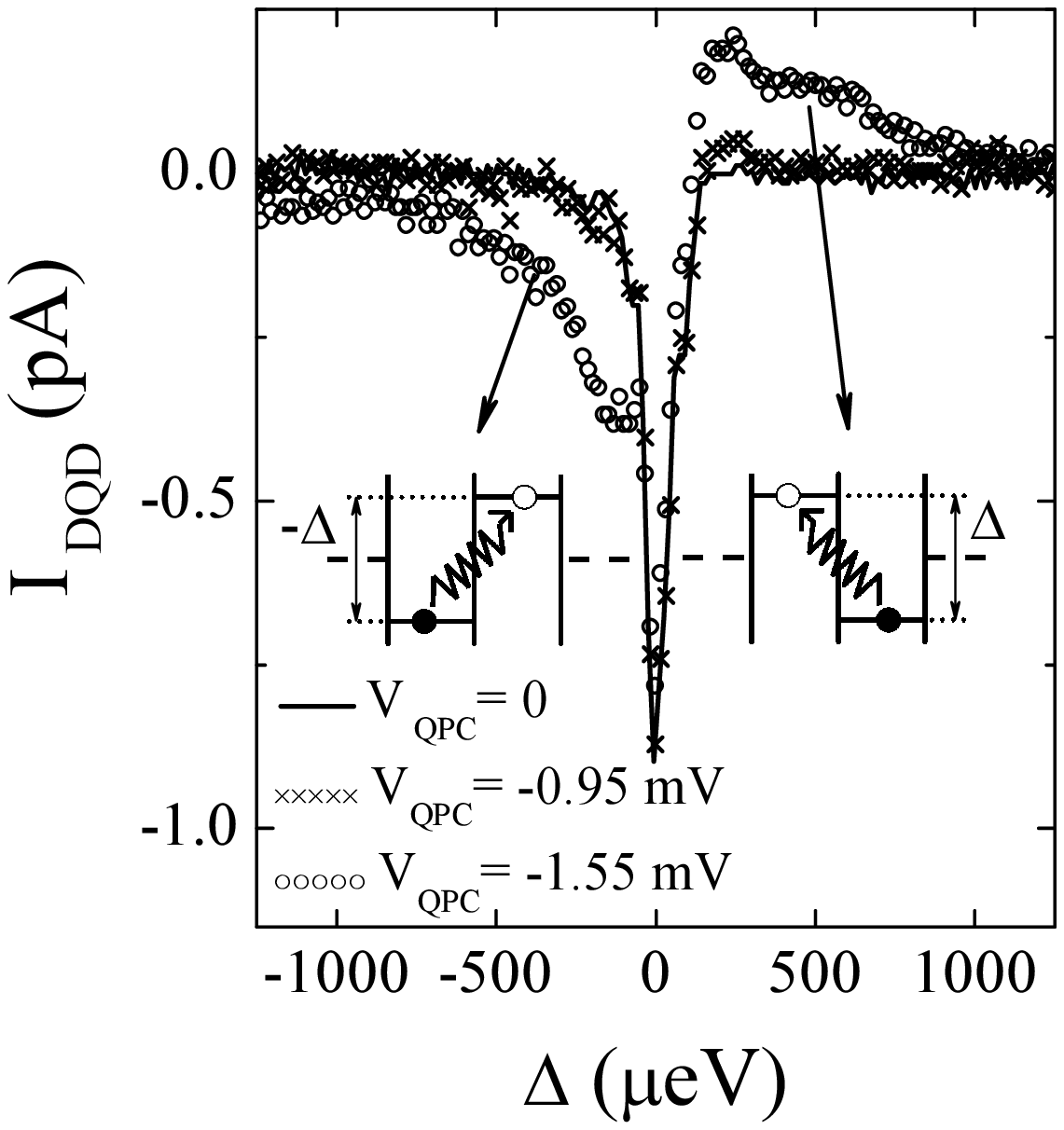} \caption{ $I_\text{DQD}(\Delta)$
taken along the dashed trace in Fig.~\ref{fig2}b. Solid line shows
the data for $V_\text{DRIVE}=0$, while crosses and circles
correspond respectively to $V_\text{DRIVE}=-0.95$~mV and -1.55~mV.
Insets: schematic inelastic tunnelling processes accompanied by
absorption of an energy quantum by the top most DQD electron.}
\label{fig3}
\end{center}
\end{figure}

The dc current generated in the DQD can be explained in terms of
energy exchange of the DQD with a strongly biased neighboring
drive-QPC. The internal asymmetry of the weakly coupled DQD in
respect to the spatial distribution of localized electron charge
makes the DQD an analog of a quantum ratchet system capable of
rectifying nonequilibrium fluctuations~\cite{reimann}. The
dependence of the generated $I_\text{DQD}$ on the position at the
charging diagram shows that the relevant charge transfer process
across the DQD quantum ratchet is an inelastic interdot tunnelling
of electrons~\cite{DQDratchet}. Under conditions of the ground
state Coulomb blockade, the interdot tunnelling requires an
absorption of an energy quantum to compensate for the lack of
energy and is useful for spectroscopic applications~\cite{aguado}.
The energy quantum should be equal to the absolute value of the
asymmetry energy, defined as the energy difference between the
[m+1,n] and [m,n+1] charge configurations ${\Delta\equiv
E_\text{[m+1,n]}-E_\text{[m,n+1]}}$.

In Fig.~\ref{fig3} we plot $I_\text{DQD}$ as a function of
$\Delta$ taken along the dashed trace in Fig.~\ref{fig2}b for
several values of $V_\text{DRIVE}$. For $V_\text{DRIVE}=0$ only
the resonant tunnelling peak at $\Delta=0$ is seen (solid line).
If $V_\text{DRIVE}$ is increased above about 1~mV, in addition to
this peak a ratchet current contribution sets in for a wide range
of asymmetry energies $|\Delta|\leq1$~meV (circles for
$V_\text{DRIVE}=-1.55$~mV in Fig.~\ref{fig3}). As expected for a
ratchet, this contribution to $I_\text{DQD}$ is asymmetric in
$\Delta$, because the inelastic interdot tunnelling of the top
most DQD electron occurs from the right to the left dot for
$\Delta>0$ and vice versa for $\Delta<0$ (see the schematics of
the absorption processes in the insets to Fig.~\ref{fig3}). The
data of Fig.~\ref{fig3} demonstrate that the drive-QPC can provide
a wide band ($\sim$250~GHz) excitation for the electrons in the
DQD quantum ratchet.

\section{Counterflow of electrons in isolated QPCs}
\begin{figure}[t]
\begin{center}
\includegraphics[clip,width=0.6\linewidth]{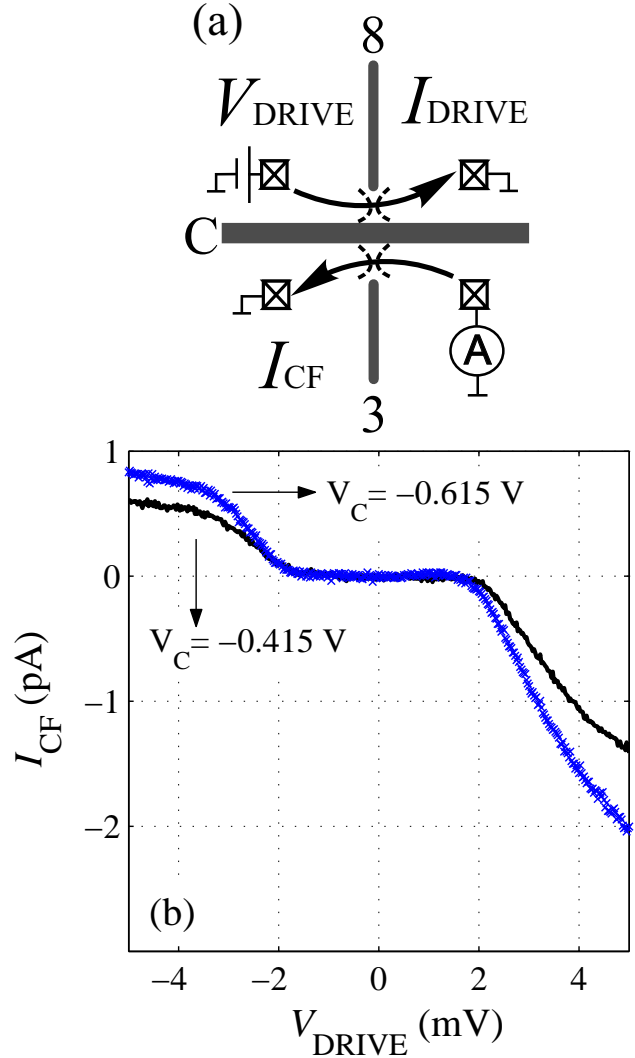} \caption{(a) -- Sketch of the counterflow measurement. The directions of
currents are shown for the case of $V_\text{DRIVE}>0$. (b) --
$I_\text{CF}$ as a function of $V_\text{DRIVE}$ for a detector-QPC
in the pinch-off regime and two values of the gate voltage
$V_\text{C}$. } \label{fig4}
\end{center}
\end{figure}

We proceed to study the mechanism of the energy transfer between
the drive and detector circuits by exchanging the detector
nanostructure by a QPC. A sketch of the measurement is shown in
Fig.~\ref{fig4}a. The detector-QPC is defined by negatively
biasing gate 3, while gates 1,2,4,5 are grounded. The detector-QPC
is tuned into the pinch-off regime, so that its lowest 1D subband
bottom is well above the leads chemical potential and the dc
conductance is very low
$G_\text{det}\simeq10~\text{G}\Omega^{-1}$. The drive-QPC
conductance is again tuned to nearly half a conductance quantum.
The source-drain voltage drop on the detector-QPC is kept at zero
and the current in the detector circuit is measured as a function
of $V_\text{DRIVE}$.

In Fig.~\ref{fig4}b we plot the current in the detector circuit
measured as $V_\text{DRIVE}$ is swept. Solid line and crosses
correspond to different gate voltages applied to the gate C,
respectively $V_\text{C}=-0.415$~V (right below the voltage above
which a detectable leakage occurs beneath the gate C) and
$V_\text{C}=-0.615$~V (strong depletion under the gate C).
Regardless of the exact value of $V_\text{C}$ substantial current
is measured in the detector-QPC at large enough $V_\text{DRIVE}$.
Remarkably, the current generated in the detector circuit is
flowing in the opposite direction to that in the drive-QPC
circuit, so we call it a counterflow current $I_\text{CF}$.

The data of Fig.~\ref{fig4}b demonstrate two important properties
of the counterflow phenomenon. First, a relatively high
$I_\text{CF}$ is detected even in a strongly depleted
detector-QPC. Second, the effect can not be suppressed by applying
a high negative voltage to gate C. The first feature signals that
the counter-flowing electrons are excited well above the Fermi
energy of the detector-QPC leads (by an energy on the order of
1~meV as DQD-ratchet data show), and hence have a much higher
probability to transmit through the nearly pinched-off
QPC~\cite{glazmann}. This observation is confirmed by an analysis
of the $I_\text{CF}$ dependence on the equilibrium transmission of
the detector-QPC~\cite{counterflowpaper}. On the other hand, the
irrelevance of the gate voltage $V_\text{C}$ rules out a direct
Coulomb interaction between the electrons of two circuits as a
possible source for energy transfer~\cite{zverev}.

\section{Conclusion}

The experiment on a DQD quantum ratchet demonstrates that a
nanostructure can be driven to a highly nonequilibrium state when
a neighboring electrically isolated drive-QPC is strongly biased.
At the same time, the observation of the counterflow phenomenon in
two QPCs indicates that the energy transfer from the drive circuit
is spatially asymmetric, i.e. the effect is caused by an energetic
imbalance across the detector-QPC in close analogy to
thermoelectric effects~\cite{molenkamp}.

Our experimental data can be qualitatively understood in terms of
an acoustic phonon based energy transfer from the drive circuit to
the detector circuit. The spatial asymmetry of acoustic phonons
emission in the drive circuit required for counterflow phenomenon
is naturally explained in the nonlinear transport
regime~\cite{counterflowpaper}. A threshold-like dependence of the
current detected in both experiments on $V_\text{DRIVE}$ might
result from a strong energy dependence of the electron-phonon
relaxation time.

The authors are grateful to V.T.~Dolgopolov, A.W.~Holleitner,
C.~Strunk, F.~Wilhelm, I.~Favero, A.V.~Khaetskii,
N.M.~Chtchelkatchev, A.A.~Shashkin, D.V.~Shovkun and P.~H\"anggi
for valuable discussions and to D.~Schr$\ddot{\text{o}}$er and
M.~Kroner for technical help. We thank the DFG via SFB 631, the
BMBF via DIP-H.2.1, the Nanosystems Initiative Munich (NIM) and
VSK the A.~von~Humboldt foundation, RFBR, RAS, and the program
"The State Support of Leading Scientific Schools" for support.

\end{document}